\documentclass[aps,prl,a4paper,twocolumn,showpacs,superscriptaddress,amssymb,amsfonts,amsmath]{revtex4}

\usepackage{bm}
\usepackage{amssymb}
\usepackage{amsfonts}
\usepackage{amsmath}
\usepackage{graphicx}
\usepackage[dvips]{color} 
\usepackage{dcolumn}
\definecolor{g-blue}{rgb}{0.83,0.95,1}
\definecolor{Blue}{rgb}{0.5,0.5,1}
\definecolor{DarkBlue}{rgb}{0.00,0.00,0.58}
\definecolor{g-yellow}{rgb}{1,1,0.7}
\definecolor{g-green}{rgb}{0.9,1,0.9}
\definecolor{green}{rgb}{0,0.6,0}
\definecolor{Green}{rgb}{0,0.4,0}
\definecolor{cyan}{rgb}{0,0.7,0.7}
\definecolor{black}{rgb}{0,0,0}
\definecolor{grey}{rgb}{0.4 ,0.4 ,0.4 }

\def\red#1{\textcolor{red}{#1}}

\begin{document}

\def \ed {\end{document}}
\def \bc {\begin{comment}} \def \ec {end{comment}}
\def\Fbox#1{\vskip1ex\hbox to 8.5cm{\hfil\fboxsep0.3cm\fbox{%
  \parbox{8.0cm}{#1}}\hfil}\vskip1ex\noindent}  


\newcommand{\eq}[1]{(\ref{#1})}
\newcommand{\Eq}[1]{Eq.~(\ref{#1})}
\newcommand{\Eqs}[1]{Eqs.~(\ref{#1})}
\newcommand{\Fig}[1]{Fig.~\ref{#1}}
\newcommand{\Figs}[1]{Figs.~\ref{#1}}
\newcommand{\Sec}[1]{Sec.~\ref{#1}}
\newcommand{\Secs}[1]{Secs.~\ref{#1}}
\newcommand{\Ref}[1]{Ref.~\cite{#1}}
\newcommand{\Refs}[1]{Refs.~\cite{#1}}

\def\be{\begin{equation}}\def\ee{\end{equation}}
\def\bea{\begin{eqnarray}}\def\eea{\end{eqnarray}}
\def\bse{\begin{subequations}}\def\ese{\end{subequations}}
\newcommand{\BE}[1]{\begin{equation}\label{#1}}
\newcommand{\BEA}[1]{\begin{eqnarray}\label{#1}}
\newcommand{\BSE}[1]{\begin{subequations}\label{#1}}

\let \nn  \nonumber  \newcommand{\br}{\\ \nn}
\newcommand{\BR}[1]{\\ \label{#1}}
\def\hf{\frac{1}{2}}
\let \= \equiv \let\*\cdot \let\~\widetilde \let\^\widehat \let\-\overline
\let\p\partial \def\pp {\perp} \def\pl {\parallel}
\def\ort#1{\^{\bf{#1}}}
\def\Trans{^{\scriptscriptstyle{\rm T}}}
\def\x{\ort x} \def\y{\ort y} \def\z{\ort z}
 \def\bn{\bm\nabla} \def\1{\bm1} \def\Tr{{\rm Tr}}
\def\Re{\mbox{  Re}}
\def\<{\left\langle}    \def\>{\right\rangle}
\def\({\left(}          \def\){\right)}
 \def \[ {\left [} \def \] {\right ]}

\renewcommand{\a}{\alpha}\renewcommand{\b}{\beta}\newcommand{\g}{\gamma}
\newcommand{\G} {\Gamma}\renewcommand{\d}{\delta}
\newcommand{\D}{\Delta}\newcommand{\e}{\epsilon}\newcommand{\ve}{\varepsilon}
\newcommand{\E}{\Epsilon}\renewcommand{\o}{\omega} \renewcommand{\O}{\Omega}
\renewcommand{\L}{\Lambda}\renewcommand{\l}{\lambda}
\renewcommand{\t}{\tau}
\def\r{\rho}\def\k{\kappa}
\def\t{\theta } \def\T{\Theta } \def\s{\sigma} \def\S{\Sigma}

\newcommand{\B}[1]{{\bm{#1}}}
\newcommand{\C}[1]{{\mathcal{#1}}}    
\newcommand{\BC}[1]{\bm{\mathcal{#1}}}
\newcommand{\F}[1]{{\mathfrak{#1}}}
\newcommand{\BF}[1]{{\bm{\F {#1}}}}

\renewcommand{\sb}[1]{_{\text {#1}}}  
\renewcommand{\sp}[1]{^{\text {#1}}}  
\newcommand{\Sp}[1]{^{^{\text {#1}}}} 
\def\Sb#1{_{\scriptscriptstyle\rm{#1}}}


\title{Turbulent Vortex Flow Responses at the AB Interface in Rotating Superfluid $^3$He-B}

\author{P.M.~Walmsley}
\affiliation{Low Temperature Laboratory, School of Science and Technology, Aalto University, FI-00076 AALTO, Finland}
\affiliation{School of Physics and Astronomy, University of Manchester, Manchester M13 9PL, UK}

\author{V.B.~Eltsov}
\affiliation{Low Temperature Laboratory, School of Science and Technology, Aalto University, FI-00076 AALTO, Finland}


\author{P.J. Heikkinen}
\affiliation{Low Temperature Laboratory, School of Science and Technology, Aalto University, FI-00076 AALTO, Finland}

\author{J.J. Hosio}
\affiliation{Low Temperature Laboratory, School of Science and Technology, Aalto University, FI-00076 AALTO, Finland}

\author{R.~H\"anninen}
\affiliation{Low Temperature Laboratory, School of Science and Technology, Aalto University, FI-00076 AALTO, Finland}

\author{M.~Krusius}
\affiliation{Low Temperature Laboratory, School of Science and Technology, Aalto University, FI-00076 AALTO, Finland}


\date{\today}

\begin{abstract}
In a rotating two-phase sample of $^3$He-B and magnetic-field stabilized $^3$He-A the large difference in mutual
friction dissipation at $0.20\,T_\mathrm{c}$ gives rise to unusual vortex flow responses. We use noninvasive NMR techniques to monitor spin down and spin up of the B-phase superfluid component to a sudden change in the rotation velocity. Compared to measurements at low field with no A-phase, where these responses are laminar in cylindrically symmetric flow, spin down with vortices extending across the AB interface is found to be faster, indicating enhanced dissipation from turbulence. Spin up in turn is slower, owing to rapid annihilation of remanent vortices before the rotation increase. As confirmed by both our NMR signal analysis and vortex filament calculations, these observations are explained by the additional force acting on the B-phase vortex ends at the AB interface.
\end{abstract}
%



\thanks{Supported by the Academy of Finland (Centers of Excellence Programme  2006-2011, grant 218211) and the EU 7th Framework Programme (FP7/2007-2013,  grant 228464 Microkelvin).}

\maketitle 

\textbf{Introduction:}--A two-phase sample of superfluid $^3$He is a unique case of coherent quantum matter,
consisting of coexisting regions of magnetic-field stabilized $^3$He-A and of $^3$He-B at zero or low field.  Here two coherent states of the same orbital $(L=1)$ and spin $(S=1)$ triplet order parameter manifold can be investigated in phase equilibrium. The structure and dynamics of vortices at the AB interface have been studied earlier in a long rotating cylinder \cite{KH}. We use this setup to examine the responses of B-phase vortices to a step change in the rotation velocity $\Omega$. At low temperatures existing vortices become easily unstable in rapid changes of rotation \cite{Precursor}. As a result, both the formation of new vortices and the annihilation of existing vortices are associated with low energy barriers. A sudden increase (reduction) in $\Omega$ leads to a continuous, but slow spin up (spin down) in the rotation of the superfluid component. In the two-phase sample at $0.20\,T_\mathrm{c}$ mutual friction dissipation is estimated to be two orders of magnitude larger in the A phase \cite{Bevan}. As a result, A-phase vortex flow is almost instantaneous while the B phase responds only slowly. This leads to unusual dynamics which differs remarkably from the laminar vortex flow observed in the absence of the A-phase stabilization field.

In a cylinder filled with superfluid $^4$He, spin up and spin down are assumed to be turbulent at essentially all temperatures below $T_\lambda$. In $^3$He-B the responses have been found to be laminar for axially homogeneous spin down or spin up at least to below $0.20\,T_\mathrm{c}$ \cite{SpinUpDownLetter,SpinUpDownJLTP}. However, in the two-phase sample B-phase spin down is faster and axially inhomogeneous. NMR line shapes, which deviate from those measured for laminar vortex flow, can here be studied in controlled conditions in the cylindrically symmetric ``flare-out" order parameter texture; in other words it is not broken cylindrical rotation symmetry, but the changed boundary conditions that cause the faster response and the increased dissipation. Thus the possibility to stabilize the A-phase layer is similar to changing in situ the boundary conditions at one of the end plates of the rotating cylinder. The resulting changes in the dynamics indicate that boundary conditions are important and provide further evidence for the fact that pinning at the end plates is responsible for the turbulent responses of superfluid $^4$He. The $^4$He vortex core diameter is of atomic size and is apparently strongly pinned at most surfaces, while in $^3$He superfluids the core diameter is at least two orders of magnitude larger.

\textbf{Experimental techniques:}--The two-phase liquid $^3$He sample is contained in a smooth-walled quartz cylinder which is mounted on a nuclear demagnetization cooling stage and can be rotated at an angular velocity $\Omega$, by rotating the cryostat (Fig.~\ref{ExpSetUp}). The time evolution and the spatial distribution of vortices is surveyed with two NMR detector coils. Two quartz tuning fork oscillators are included for temperature measurement. A small  superconducting solenoid around the cylinder provides the axially oriented magnetic field for stabilizing an A-phase layer which divides the NMR sample in two identical B-phase sections.

\begin{figure}[t]
\begin{center}
\centerline{\includegraphics[width=0.95\linewidth]{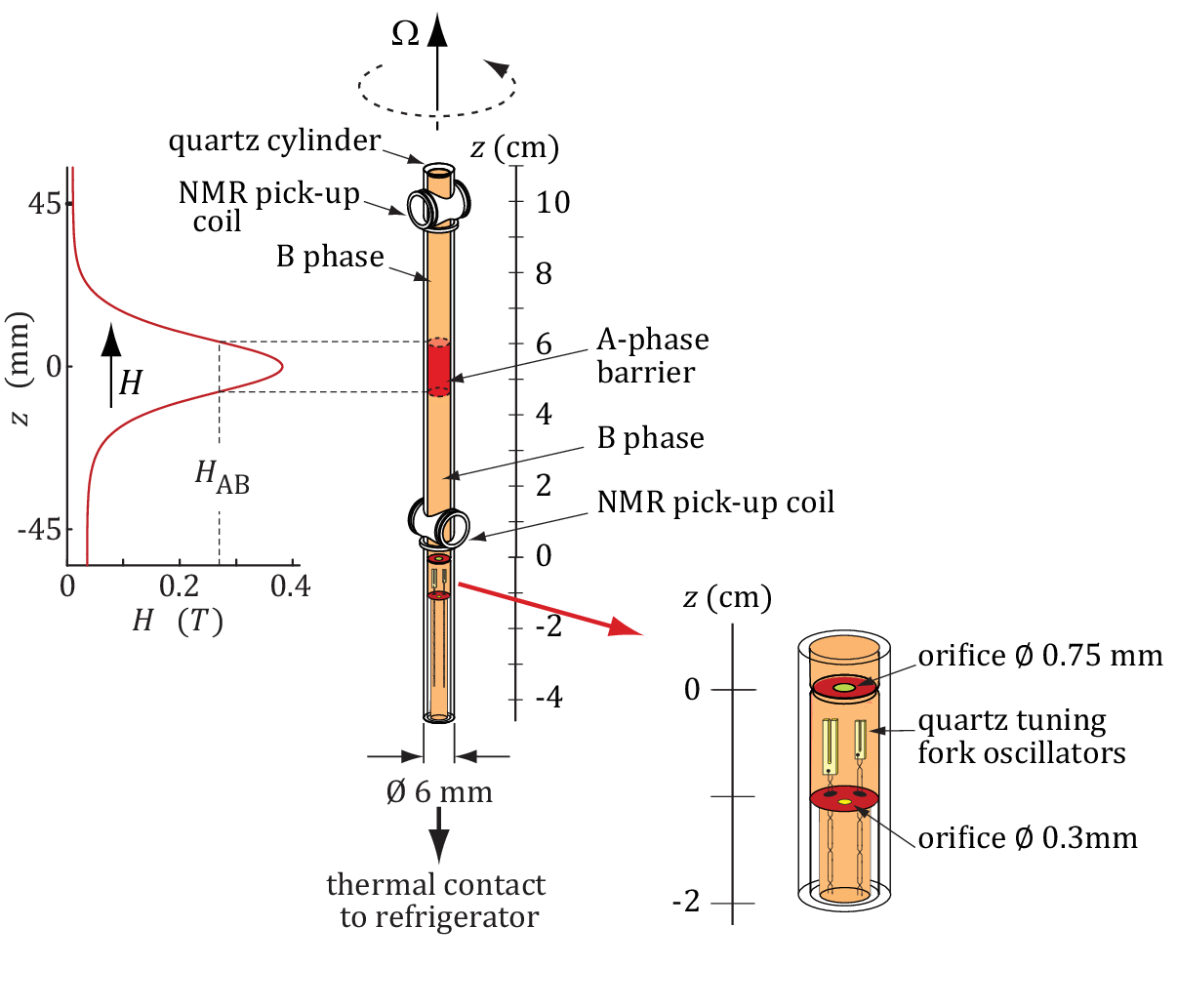}}
\vspace{-5mm}
\caption{Measuring setup. Two independent cw NMR spectrometers monitor vortex motion in the 110\,mm long section of the sample cylinder above the $\varnothing$\,0.75\,mm orifice. The middle section with the quartz tuning fork oscillators is employed for thermometry. The bottom section below the $\varnothing$\,0.3\,mm orifice provides thermal contact to the refrigerator. The superconducting solenoid in the center of the NMR sample produces the axially oriented barrier field $H_\mathrm{b}$ for stabilizing $^3$He-A. With a current of 6.2\,A in the barrier magnet the width of the A-phase layer is 10\,mm.
}
\label{ExpSetUp}
\end{center}
\vspace{-10mm}
\end{figure}

Vortex structures in the A and B phases are incommensurable, but they do interconnect across the AB interface after the doubly-quantized A-phase vortices dissociate and  terminate in point defects on the phase boundary \cite{AB-Interface}. The critical velocity of vortex formation is low in the A-phase section \cite{Kopu} and in rotation it is approximately in the equilibrium vortex state, where both the normal and superfluid components are in solid-body rotation with the container, ${\bm v}_{\rm n} =\bm{\Omega}\times{\bm r} \approx {\bm v}_{\rm s}$. In the B-phase sections the critical velocity is an order of magnitude higher \cite{B-CritVel}. Established procedures exist for maintaining in the B-phase sections either vortex-free counterflow, where the superfluid fraction is at rest in the laboratory frame (${\bm v}_{\rm s} = 0$), or a flow state with a central vortex cluster with a known number of vortex lines $N$. The cluster is formed from rectilinear vortices at the solid-body-rotation density $2\Omega/\kappa$, so that within the cluster ${\bm v}_{\rm s}  \approx {\bm v}_{\rm n}$ while outside ${\bm v}_{\rm s} = \kappa \, N/(2\pi r)\,\hat{\bm e}_\phi$ ($\kappa = \hbar/(2 m_3)$ is the $^3$He circulation quantum). Thus there exist different configurations from where spin-down/spin-up measurements can be started. These can be classified according to how the $\Omega$ range of the measurement interacts with the AB interface instability, which usually is characterized by its critical angular velocity $\Omega_\mathrm{AB}(T,P)$ measured in a situation when the A phase is in the equilibrium vortex state and the B phase is vortex-free \cite{ROP}.

Our spin-down measurement starts from the equilibrium vortex state (at $\Omega_\mathrm{i}$) which extends through all three sample sections, with vortices interconnected across the two AB interfaces. Choosing $\Omega_\mathrm{i} < \Omega_\mathrm{AB}(T,P)$, the AB interface instability plays no role in these results. With decreasing temperature the $^3$He-B mutual friction dissipation $\alpha (T)$ drops increasingly below that of $^3$He-A and its characteristic dynamic response time, $\tau (T) \propto 1/\alpha (T)$, is slowed down.  The faster A-phase dynamics provides then additional pull on the B-phase vortex ends on the phase boundary. This leads to faster and axially inhomogeneous B-phase spin down (Fig.~\ref{VortexConfiguration}). The ensuing events are the following: (i) The A-phase vortices spiral rapidly in laminar motion to the cylindrical wall. This exerts a force on the B-phase vortices such that the vortex density in the center of the cylinder is depleted faster than at larger radii $r$ and at larger distances $\Delta z$ from the AB interface. The central depletion in vortex density proceeds to such extent that it can be characterized as a vortex-free dome on the AB interface. Thus the vortex density is not constant across the cross section of the cylinder and the superfluid fraction is not in solid-body rotation, but relaxes to the final state ${\bm v}_{\rm s} = 0$ faster in the center than at large radii. (ii) Because of the central vortex-free dome, which borders to the AB interface and decreases in radius further away at increasing $z$, the vortices become curved along the cylinder axis $z$. Differences in the azimuthal velocities as a function of $z$ cause them to become helically twisted \cite{TwistedCluster}. The inhomogeneous twist introduces reconnections and turbulence, which speed up the B-phase response. On the AB interface the remaining vortices curve parallel to the boundary and extend radially outward, ending perpendicularly on the cylinder wall. This is a B-phase vortex sheet, analogous to the A-phase vortex sheet covering the AB interface in rotation when the B-phase is free of vortices \cite{AB-Interface}.

\begin{figure}[t]
\begin{center}
\centerline{\includegraphics[width=0.95\linewidth]{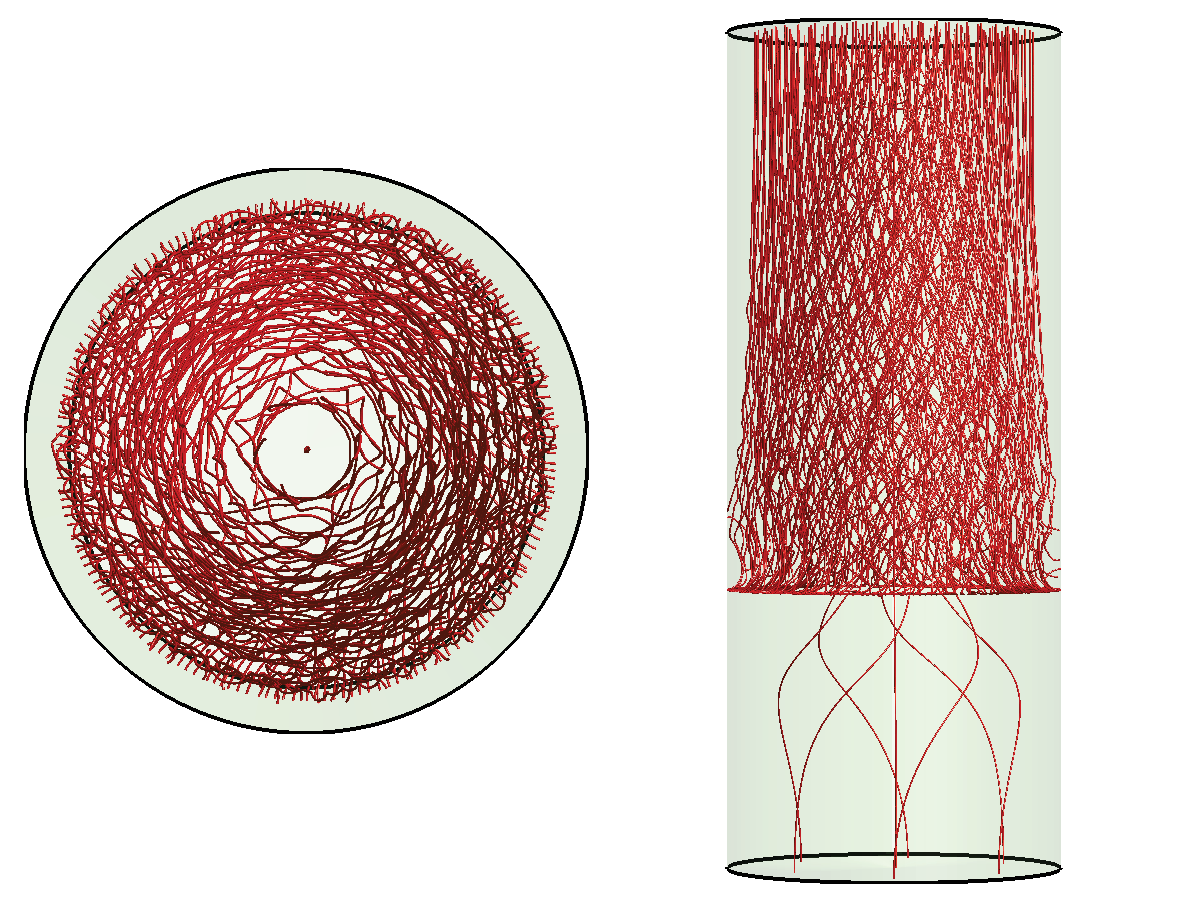}}
\vspace{-5mm}
\caption{Calculated vortex configuration during spin down of the superfluid component, 30\,s after a step-like reduction of $\Omega$ from 0.25\,rad/s to zero. \textbf{\textit{(Left)}} View from the top into the cylinder showing the empty hollow in the center and twisted vortices around it. \textbf{\textit{(Right)}} Side view showing the A-phase section on the bottom, with only a few vortices left in the outermost ring, while in the B-phase the AB interface at large $r$ is covered with a vortex sheet \cite{AB-Interface}. Above the interface the vortices are helically twisted around the central axis. Their polarization parallel to the axis increases with distance from the interface. Parameters: $T = 0.22\,T_\mathrm{c}$ and $P = 29\,$bar, which corresponds to $\alpha_\mathrm{A} \approx 2$ and $\alpha_\mathrm{B} = 4.3 \times 10^{-3}$ [while $\alpha^\prime_\mathrm{A} \approx 0.8$ and $\alpha^\prime_\mathrm{B} \approx 0$], cylinder radius $R = 3\,$mm, the lengths of the A and B-phase sections are 5 and 10\,mm, the cylinder axis is aligned parallel to the rotation axis.
}
\label{VortexConfiguration}
\end{center}
\vspace{-10mm}
\end{figure}

Spin up, in turn, depends on the number and configuration of remanent vortices in the B-phase section. The influence of the AB interface is to speed up the removal of remnants. Thus also spin up is different from that measured for the single-phase sample in the same conditions. These differences in the spin-down/spin-up responses of the two-phase and single-phase samples are most noticeable at the lowest temperatures. We concentrate here on measurements at 29\,bar liquid $^3$He pressure and $0.20\,T_\mathrm{c}$. This is the minimum temperature for the current setup, which is limited by a residual heat leak of $\sim 15$\,pW through the lower orifice of 0.3\,mm diameter.

\begin{figure}[t]
\begin{center}
\centerline{\includegraphics[width=0.99\linewidth]{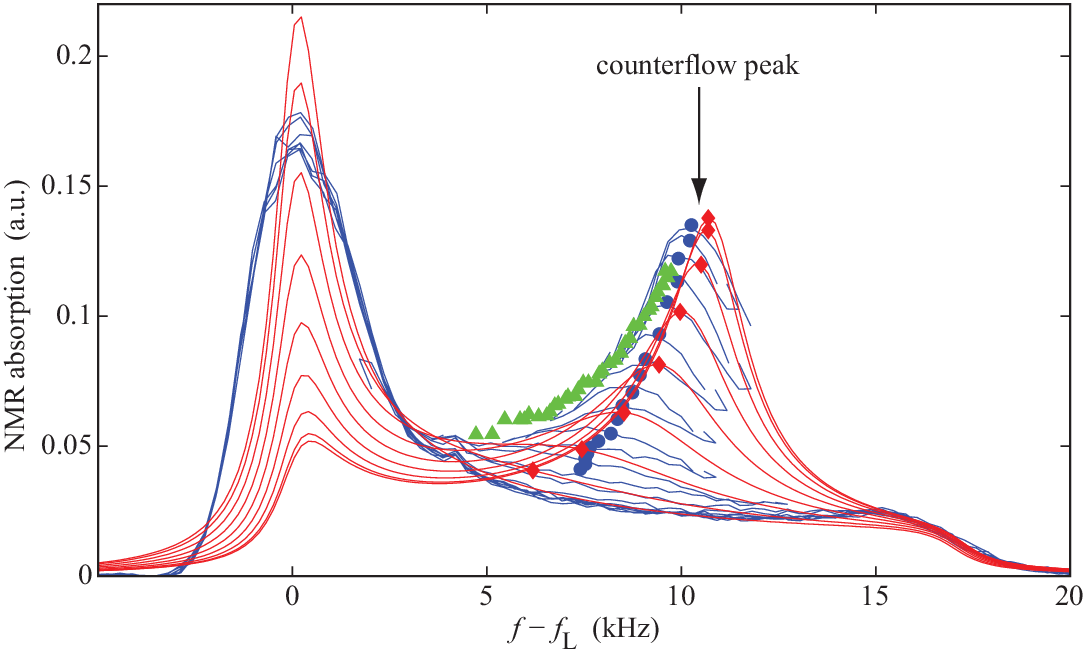}}
\caption{NMR absorption spectra during spin down. The line shapes \textit{(red curves)} have been calculated using the cylindrical shell model for the two-phase sample and show the counterflow (cf) peak on the right at large frequency shifts $f - f_\mathrm{L}$ from the Larmor frequency $f_\mathrm{L}$. The decreasing cf peak heights \textit{(red lozenges)} correspond to increasing radius $r_\mathrm{s}$ of the central vortex-free cylinder in $20\,\mu$m steps, starting from $r_\mathrm{s} = 0$. The \textit{blue curves} are from the measurements where the cf peak is monitored continuously by sweeping around its maximum \textit{(blue dots)} during its decay. The \textit{green triangles} show the measured peak trajectory in laminar spin down in the absence of the A-phase layer \cite{deGraaf}. Parameters: $T = 0.20\,T_\mathrm{c}$, $P = 29\,$bar, $\Omega_\mathrm{i} = 1.0\,$rad/s, $d\Omega/dt = -0.03\,$rad/s$^2$.
}
\label{CF-Peak}
\end{center}
\vspace{-10mm}
\end{figure}

\textbf{NMR techniques:}--In vortex-free rotation at constant $\Omega$ the B-phase order parameter texture is modified by the azimuthally flowing superfluid counterflow (cf) at the velocity $\bm{v} = \bm{v}_\mathrm{n} - \bm{v}_\mathrm{s} = \bm{\Omega} \times \bm{r}$. This solid-body-like cf reorients the B-phase anisotropy axis, which is induced by the NMR polarization field. The order parameter distribution is thereby changed, and a large NMR frequency shift appears. In the NMR spectrum the shift is expressed as a so-called cf peak which can be calibrated as a function of the experimental variables $\Omega$, $T$, and $P$ \cite{deGraaf}. The result is a unique trajectory as a function of $\Omega$ for the peak absorption \textit{vs} the frequency shift at fixed $T$ and $P$, which identifies the solid-body velocity distribution (Fig.~\ref{CF-Peak}). During the slow spin down of the superfluid fraction after a sudden stop of rotation, when $\bm{v}_\mathrm{n} = 0$ and $\bm{v}_\mathrm{s}$ arises from the distribution of vorticity, also a cf peak is formed. If the spin-down response is laminar, as is the case for the single-phase sample, the vortices remain approximately straight and maintain a constant density across the cross section of the cylinder, \textit{i.e.} they are in the solid-body configuration and the trajectory of this transient cf peak matches that measured in the vortex-free state at stationary conditions as a function of $\Omega$ \cite{SpinUpDownLetter,SpinUpDownJLTP}.

The cf peak recorded during the spin down of the two-phase sample follows a different trajectory of peak height \textit{vs} frequency (Fig.~\ref{CF-Peak}). Its frequency shift proves to be larger, presumably owing to the compression of vortices to large radii around the vortex-free dome in the center. This observation suggests a simple model how to analyze the changed vortex distribution. Assume that all vortices are compressed into an outer cylindrical shell with the initial vortex density $2 \Omega_0 /\kappa$ and that the empty inner region increases gradually in radius $r_\mathrm{s}$. In this model the azimuthal velocity of the superfluid component is $\bm{v}_\mathrm{s} \equiv 0$ inside the central cylinder $r < r_\mathrm{s}$, while in the outer cylindrical shell $r_\mathrm{s} < r < R$ it is $v_\mathrm{s} \approx \Omega_0 (r^2 - r_\mathrm{s}^2)/r$. Here $\Omega (t = 0) = \Omega_0  \approx 1.0\,$rad/s is the value extrapolated back to $t = 0$, the moment when the cryostat rotation comes to a stop and the cf peak height passes through its maximum value. To compare results with and without the A-phase layer, we use the normalized solid-body vortex density equivalent $\Omega / \Omega_0 = 1 - (r_\mathrm{s} / R)^2$. In Fig.~\ref{CF-Peak} the NMR spectra have been calculated for every $20\,\mu$m step increase in $r_\mathrm{s}$ \cite{TextureProgramme}. The trajectory of the cf peaks calculated in this way agrees closely with that of the measured spin-down response. Thus the simple model with only azimuthal cf appears to capture the dominant features, although it neglects all dependence on the distance from the AB interface and on twisted or tangled vortices.

\begin{figure}[t]
\begin{center}
\centerline{\includegraphics[width=1\linewidth]{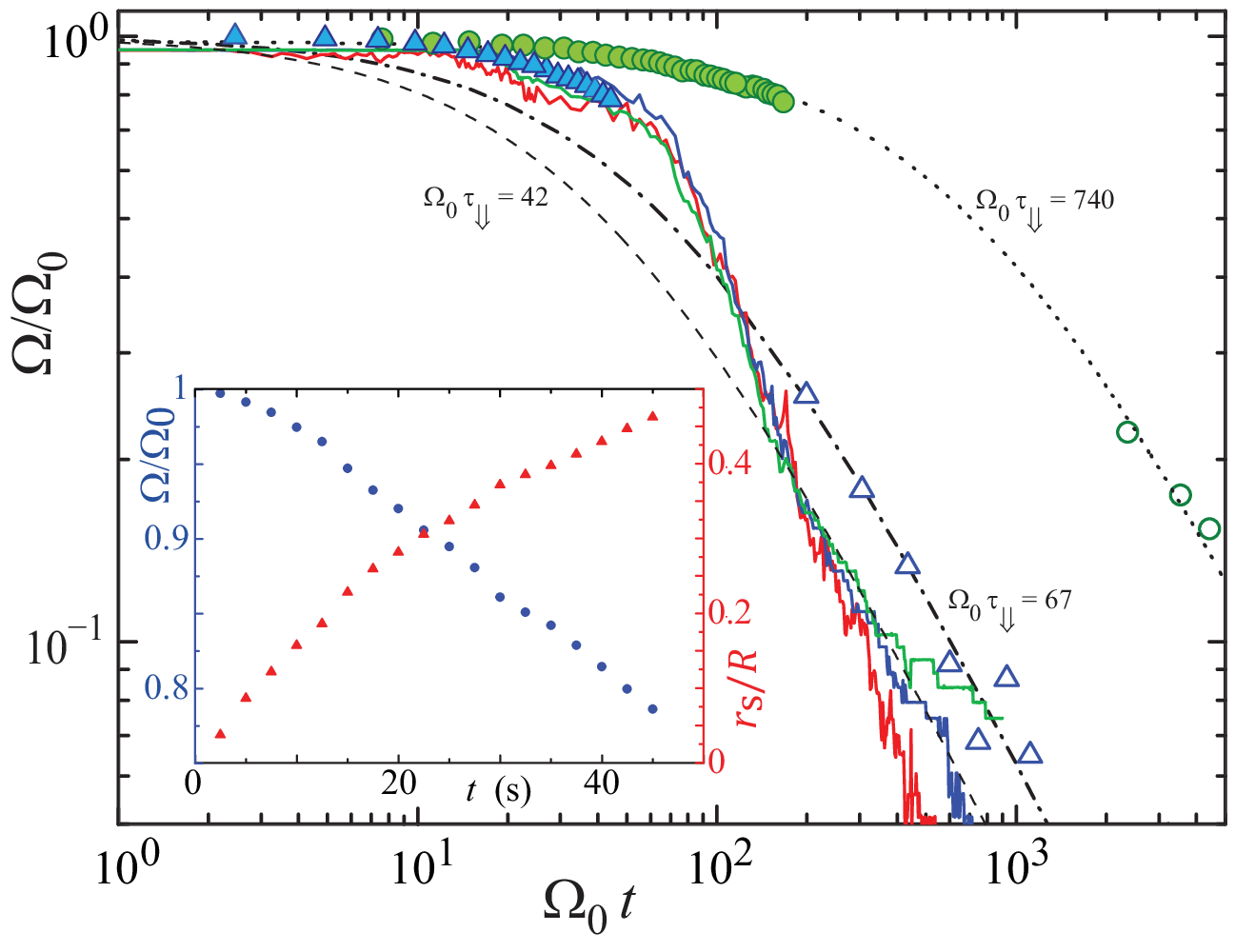}}
\vspace{-5mm}
\caption{Normalized spin-down of the azimuthal flow $\Omega (t) / \Omega_0$. The response of the two-phase sample \textit{(blue triangles)}, analyzed as shown in the inset, is compared to that measured in the absence of the A phase layer \textit{(green circles)}. \textit{Filled symbols} correspond to cf peaks measured during spin down, while \textit{open symbols} represent extrapolations from subsequent spin ups (see text). The \textit{broken line} curves represent laminar fits with $\Omega (t) = \Omega_0 / (1+t/\tau_\Downarrow)$, where $\tau_\Downarrow = (2 \alpha \Omega_0)^{-1}$: the \textit{dotted} curve with  $\tau_\Downarrow = 740 \,$s represents the fit to the data measured in the absence of the A-phase layer, while the \textit{dash-dotted} curve is a fit to the laminar tail of the two-phase-sample data with $\tau_\Downarrow = 67 \,$s. The \textit{solid} curves are calculations of the normalized azimuthal velocity $v_{\mathrm{s},\phi} (R) /(\Omega_0 R)$ with the parameter values from Fig.~\ref{VortexConfiguration}. The three curves represent different distances $\Delta z$ from the AB interface: red, $\Delta z = 1\,$mm; blue, $\Delta z = 5\,$mm; green, $\Delta z = 10\,$mm. The \textit{dashed} fit with $\tau_\Downarrow = 170 \,$s is the laminar tail of the late-time average of the three calculated curves.  \textit{\textbf{(Inset)}} Fit to the cylindrical shell model of the two-phase sample. \textit{(Right vertical axis)} Normalized radius $r_\mathrm{s} / R$ (red triangles) of the vortex-free central cylinder, when fitted to the measured spin-down as a function of time $t$. \textit{(Left vertical axis)} Equivalent normalized solid-body vortex density $\Omega / \Omega_0 = 1 - (r_\mathrm{s} / R)^2$ (blue dots).
}
\label{SpinDownMeasuredResultAB}
\end{center}
\vspace{-10mm}
\end{figure}

\textbf{Spin-down measurements:}--For quantitative analysis, the decreasing cf peaks at $t \geq 0$ were fitted by means of the texture calculation procedure to the cylindrical shell model with one free parameter, by adjusting the inner cylinder radius $r_\mathrm{s}$ and assuming a vortex density outside, $2 \Omega_0 /\kappa$, which is constant as a function of $r$ and $t$. The resulting fit in the inset of Fig.~\ref{SpinDownMeasuredResultAB} reveals a rapidly growing central region where the superfluid component has already come to rest: in 50\,s the radius of the vortex-free central region has grown to half of the cylinder radius and at this point $\sim 25$\,\% of the vorticity has annihilated.

In Fig.~\ref{SpinDownMeasuredResultAB} the spin-down responses of the azimuthal flow $\Omega (t) / \Omega_0$ are compared in the two cases, with and without the A-phase layer. The measurement is performed by decelerating the rotation drive from $\Omega_\mathrm{i}$ to zero at $d\Omega/dt = - 0.03\,$rad/s$^2$ and the cf peak is recorded as a function of time. With decreasing temperature the cf peak measurement procedure runs into problems: the cf peak height is reduced and reaches zero at higher vortex-free flow velocities \cite{deGraaf}. At $0.20\,T_\mathrm{c}$ in Fig.~\ref{SpinDownMeasuredResultAB} the cf peak is lost when the vorticity has dropped by $\sim 25$\,\% \textit{(filled symbols)}.  This explains the blank region where there are no data points. In this region more data \textit{(open symbols)} can be retrieved by increasing rotation suddenly back to some large value, where the cf peak can be recorded while it decays during spin up. This data can then be extrapolated back to the moment when rotation was increased. Combining the two data sets, the spin-down response is seen to be a smooth monotonic decay, both with and without the A-phase layer. However, in the former case it is appreciably faster and not of the laminar form $\Omega (t) = \Omega_0 / (1+t/\tau_\Downarrow)$, where a single time constant $\tau_\Downarrow = (2 \alpha \Omega_0)^{-1}$ fits the data.

For comparison, Fig.~\ref{SpinDownMeasuredResultAB} also shows the corresponding spin-down response for the calculation in Fig.~\ref{VortexConfiguration} at different distances $\Delta z$ from the AB interface. We find that the result is not very sensitive on $\Delta z$. Comparing the shapes of the response curves we see that in the presence of the A-phase layer an abrupt shoulder appears, resulting from faster spin down than dictated by the laminar dependence. This is then towards the end of the decay followed by a slow final laminar tail. The fast section of spin down decay is present in both the calculated and measured results in the presence of the A-phase layer. This difference from the laminar dependence quantifies the turbulent dissipation \cite{Walmsley}.

\begin{figure}[t]
\begin{center}
\centerline{\includegraphics[width=0.95\linewidth]{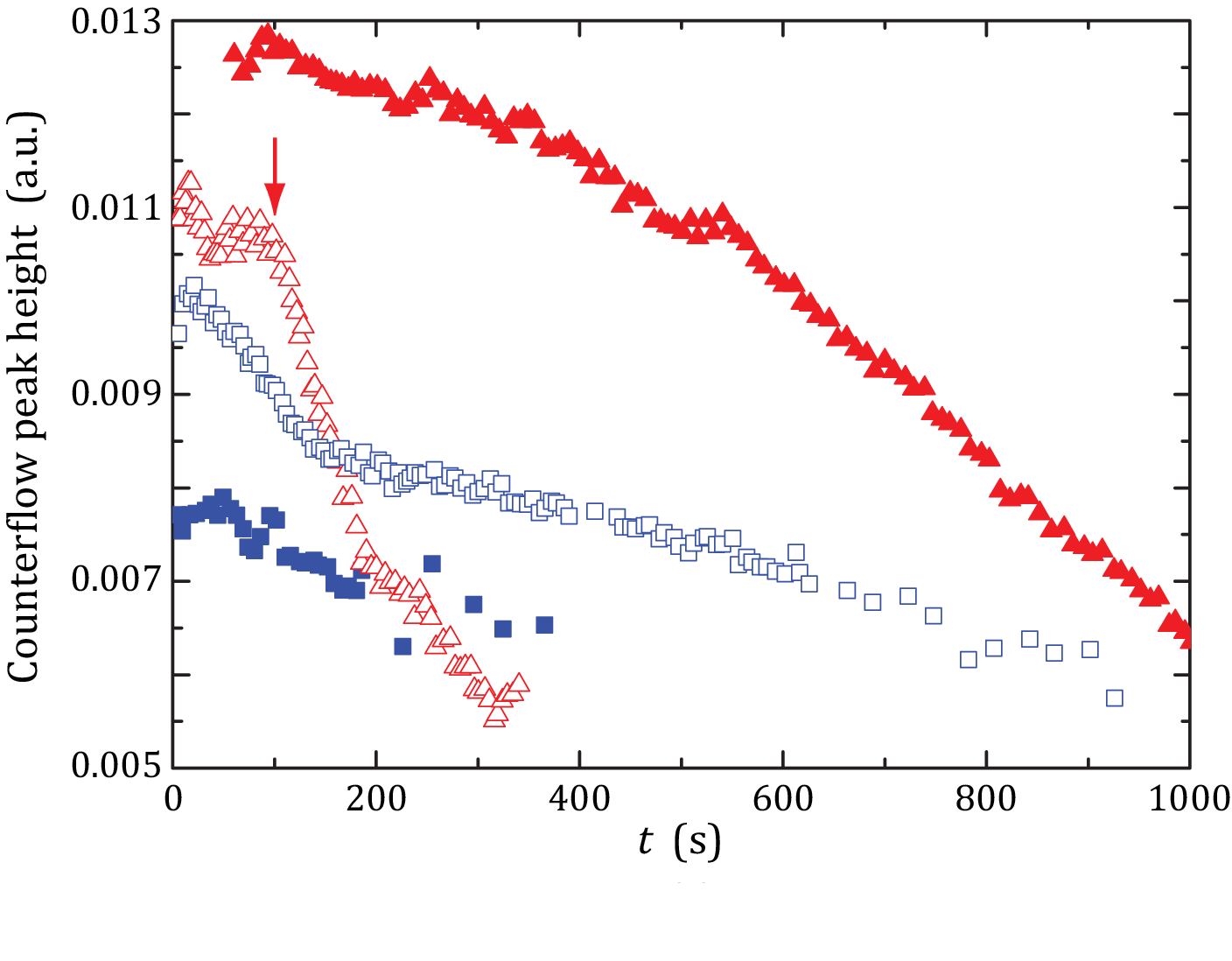}}
\vspace{-7mm}
\caption{Measured cf peak height as a function of time in spin up. \textit{(Filled symbols)} Spin up after having waited for 40\,min at $0.20\,T_\mathrm{c}$ in zero rotation for remanent vortices from a previous spin-down measurement to annihilate. The response is much slower in the presence of the A-phase layer \textit{(red triangles)} than in its absence \textit{(blue squares)}. \textit{(Open symbols)} Spin up from a state with $\sim 120$ remnants. Spin up is now faster with the A-phase layer \textit{(red triangles)} than without \textit{(blue squares)}. The vertical arrow denotes the moment when the vortex front enters the NMR coil. In these measurements rotation is increased from zero to $\Omega_\mathrm{f} = 1.0\,$rad/s at $d\Omega/dt = 0.03\,$rad/s$^2$. Time $t = 0$ marks the moment when $\Omega_\mathrm{f}$ is reached. The number of remnants is estimated from the cf peak height extrapolated to $t = 0$.
}
\label{SpinUp}
\end{center}
\vspace{-10mm}
\end{figure}

\textbf{Spin-up measurements:}--In similar manner we can determine from the cf peak response the decay of azimuthal flow during spin up, after a rapid increase of $\Omega$ \cite{SpinUpDownJLTP}. We restrict the discussion to the simplest case where the acceleration is started from rest ($\Omega_\mathrm{i} = 0$) and is finished at $\Omega_\mathrm{f} < \Omega_\mathrm{AB}(T,P)$. In the A-phase section vortices are formed rapidly and independently during the rotation increase. On the AB interfaces they curve radially outward, covering the interface as a vortex sheet \cite{AB-Interface}. Each of these connects to a B-phase vortex across the AB interface later, when B-phase vortices are formed. The B-phase spin up may proceed in two different ways, depending on the number and configuration of remanent vortices in the initial state at rest \cite{Remnants}. (1) If there is a fair number of remnants evenly distributed along the B-phase section, then the response might be a slow axially homogeneous build up of the vortex density with solid-body distribution, which is carried to completion all the way to the equilibrium vortex state. In this case the cf peak follows the solid-body rotation trajectory in Fig.~\ref{CF-Peak}. (2) If there are only few remnants, the slow axially homogeneous response might be terminated in a localized turbulent burst of vortex formation, followed by a subsequent axial expansion of the vortices along the rotating cylinder \cite{PLTP}. This spin-up process results in a markedly different cf peak response, one that depends on the distance of the turbulent burst from the detector coil.

In Fig.~\ref{SpinUp} two cases of spin up with and without the A-phase layer are compared. Here the remnants are left-over vortices from a previous spin-down measurement. In one case a waiting period of 40\,min is enforced at zero rotation, to allow for annihilation. This is calculated from the moment when rotation comes to a stop after the preceding spin-down measurement. Owing to the faster spin-down in the presence of the A-phase layer, the number and length of remnants is smaller so that the subsequent spin-up is prominently slower \textit{(red filled triangles)} than in the absence of the interface \textit{(blue filled squares)}. In contrast, when the number of remnants is adjusted to be roughly equal and relatively large, the spin-up response becomes faster in the presence of the A-phase layer \textit{(red open triangles)}, owing to a turbulent burst of vortex formation at the AB interface which starts the axial motion of vortices along the cylinder. The passage of the vortex front through the NMR coil is indicated by an abrupt decline of the cf peak height \cite{Front}.  Presumably the turbulent burst is triggered by the complex reconnection processes which take place in the vicinity of the AB interface when A and B phase vortices connect across the interface and move at different velocities.

\textbf{Spin-down calculations:}--Vortex filament calculations have proven instructive for analyzing spin-down responses, while vortex formation has turned out to be problematic in spin-up \cite{PLTP}. Our calculation of spin down, of which Fig.~\ref{VortexConfiguration} is a snapshot, makes the following two simplifications: all vortices are considered to be similar and singly quantized, while the AB interface is treated as a plane where mutual friction dissipation changes discontinuously from $\alpha_\mathrm{A}$ to $\alpha_\mathrm{B}$ at $z = 5\,$mm. The viscous normal component is locked to the rotation drive. At low rotation $\Omega_\mathrm{i} < \Omega_\mathrm{AB}$ the calculation describes the dynamics properly, in spite of the simplifications, and reproduces the rapid response in the A-phase section, the formation of a B-phase vortex sheet on the AB interface, and the helically twisted vortices in the B phase, with all the currents which follow from these configurations \cite{URL}.

\begin{figure}[t]
\begin{center}
\centerline{\includegraphics[width=1.1\linewidth]{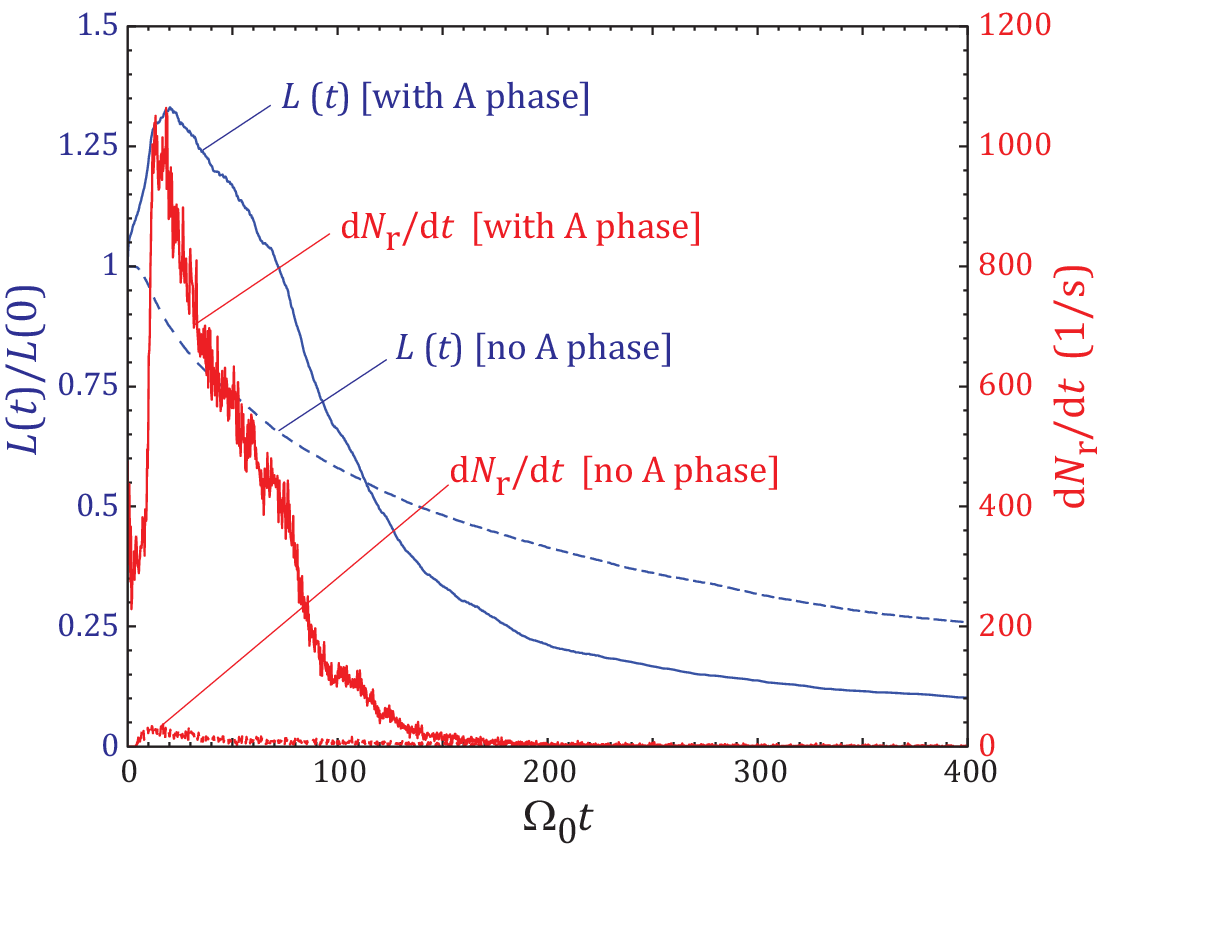}}
\vspace{-10mm}
\caption{Calculated B-phase spin down for the example in Fig.~\ref{VortexConfiguration}, emphasizing differences in the response with and without the A-phase section. \textit{(Red)} Reconnection rate $dN_\mathrm{r} / dt$ as a function of normalized time $\Omega_0 t$ \textit{(solid line)}. The non-monotonic rate peaks sharply at around 70\,s and then declines rapidly in $\sim 400\,$s to 10\,\% of the peak value. In comparison, the reconnection rate is orders of magnitude smaller and limited to a surface layer in the absence of the A-phase section \textit{(dashed line)}. \textit{(Blue)} The normalized total vortex length $ L (t)/L(0) $ also displays an overshoot and peaks at around 90\,s \textit{(solid line)}. It then declines rapidly in 900\,s to 10\,\% of its peak value. The later decay is slower with $t^{-1}$ dependence. In the absence of the A-phase section $ L (t)/L(0) $ displays a monotonically decreasing curve with slow $(1+t/\tau_\Downarrow)^{-1}$ dependence \textit{(dashed line)}. The parameters are the same as in Fig.~\ref{VortexConfiguration}. The reference curves with no A-phase section have been calculated for a cylinder of 10\,mm length, tilted by $\eta = 2^\circ$ from the rotation axis, to break cylindrical symmetry.
}
\label{SpinDownCharacteristics}
\end{center}
\vspace{-10mm}
\end{figure}

Fig.~\ref{SpinDownCharacteristics} summarizes some of the differences in the spin down with and without the A-phase section. Most important is the much larger frequency of vortex reconnections \textit{(red curves)} and their spatial distribution. Since inter-vortex reconnections in the bulk volume feed the turbulence, it is instructive to compare them in the two cases. Roughly half of all reconnections occurs within 0.2\,mm of the AB interface while the other half is distributed relatively evenly as a function of $z$ above the AB interface. Radially the reconnections increase steeply towards large radii close to the AB interface, while away from the interface the increase is slower. In contrast, in the absence of the A phase section spin down is laminar and the reconnections are concentrated to a narrow surface layer on the cylindrical wall, leaving the central bulk volume reconnection free \cite{SpinUpDownJLTP}. Thus here we have a clear difference -- in the former case reconnections occur in the bulk volume and on the AB interface, in particular, while the surface layer is in the latter case the place where the annihilating and reconnecting vortices transfer their angular momentum to the walls.

The comparison on the decay of the total vortex length $L (t)$ underlines in a similar manner the differences between turbulent and laminar responses \textit{(blue curves)}: turbulent spin down in the presence of the A-phase section is non-monotonic with an initial overshoot and a subsequent rapid decay, which might be characterized with a $t^{-3/2}$ dependence \cite{Walmsley}. Here part of the kinetic energy of the superfluid component is dissipated in reconnections and other turbulent excitations. Subsequently the decay turns into the slower $t^{-1}$ dependence in the laminar tail. In contrast, in the absence of the A phase section $L (t)$ decays monotonically with the laminar $(1+t/\tau_\Downarrow)^{-1}$ dependence.  Overall we find close correspondence between the calculated and measured spin-down responses which indicates that differences in vortex velocities across the AB interface explain adequately our results.

\textbf{Conclusions:}--The two-phase superfluid $^3$He sample provides a unique environment for vortex studies in the zero temperature limit, with two interacting superfluids and their vastly different time scales of vortex flow. This leads to unusual vortex configurations in the B-phase section during spin down: a vortex-free dome is formed in the center, surrounded by a vortex sheet on the AB interface and an outer shell of helically twisted vortices at high density. Reconnections among the twisted vortices and on the AB interface concentrate to large radii. They bring about increased dissipation and a faster turbulent spin-down response. These differences from axially homogeneous laminar spin down in the absence of the A-phase make it possible to identify the first changes in the response from fully laminar towards weakly turbulent, which arise from the influence of a well-defined planar perturbation. The resulting spin-down response is different from that when the perturbation is present at all surfaces, such as the spin down of a viscous fluid in a cylinder \cite{Greenspan}, or the spin down of a cube filled with superfluid He \cite{Walmsley,SpinUpDownJLTP}.

This work is supported by the Academy of Finland (Centers of Excellence Programme  2006-2011, grant 218211) and the EU 7th Framework Programme (FP7/2007-2013,  grant 228464 Microkelvin).

\end{document}